\documentclass[ twocolumn,aps,prd,   
               preprintnumbers,numbers,sort&compress,
               nofootinbib,
                            showpacs,
               colorlinks,
               linkcolor=blue,
               citecolor=blue]{revtex4-1}
   \newcommand{\exclude}[1]{}

\usepackage{graphicx,amsmath,amssymb,bm}
\usepackage{psfrag}
 \usepackage{feynmp}
\usepackage{hyperref}
\usepackage{enumitem}

\providecommand{\d}{}
\renewcommand{\d}{{\rm d}}

\newcommand{\beq}{\begin{equation}}
\newcommand{\eeq}{\end{equation}}
\newcommand{\be}{\begin{eqnarray}}
\newcommand{\ee}{\end{eqnarray}}

\newcommand{\GeV}{\text{ GeV}}
\newcommand{\MeV}{\text{ MeV}}

\begin{document} 

\title{Quark Nugget Dark Matter: No contradictions with neutrino flux constraints.} 
\author{K. Lawson and A. R. Zhitnitsky}
\affiliation{ Department of Physics \& Astronomy, University of British Columbia, 
Vancouver, B.C. V6T 1Z1, Canada} 

\begin{abstract}
It has recently been claimed \cite{Gorham:2015rfa} that dark matter in the form of quark nuggets 
cannot account for more than 20$\%$ of the dark matter density. This claim is based on 
constraints on the neutrino flux in the 20-50 MeV range where the sensitivity of underground 
neutrino detectors such as Super-Kamiokande have their highest signal-to-noise ratio. 
We stress the dependence of this claim on the details of the annihilation 
of visible baryons with an antiquark nugget and of the resulting neutrino spectrum. 
In particular if the annihilation occurs in the bulk of an antiquark 
nugget in the colour superconducting (CS)  
phase the lightest pseudo Goldstone mesons (the pions and Kaons) have masses in the 
20 MeV range, in contrast with the conventional mesons for which 
$m_{\pi}\simeq 140$ MeV. Thus, the decay of 
these light pseudo Goldstone bosons of the CS phase cannot produce highly energetic neutrinos 
in the 20-50 MeV energy range. We predict the basic properties of the resulting neutrino 
spectrum and argue that, while nuggets composed of purely nuclear matter would be effectively 
ruled out, there remains a wide class of CS models which are fully consistent with   
present neutrino flux constraints. However, future neutrino observations may offer a 
strong potential test of these models. 
\end{abstract}
\maketitle

\section{Introduction} 
This work is motivated by the recent preprint \cite{Gorham:2015rfa}, in which the 
authors claim that the interactions of (anti) nuggets with normal matter in the Earth and 
Sun will lead to annihilation and an associated neutrino flux. 
As the antineutrino spectrum is strongly constrained by SuperK observations it was 
argued in \cite{Gorham:2015rfa} that nuggets producing a significant neutrino flux 
in the 20-50 MeV range cannot account for more than 20$\%$ of the dark matter 
density. In response to this claim we conduct a more extensive analysis of the neutrino 
spectrum than has been previously provided and demonstrate that the spectrum is 
in fact peaked at lower energies where the constraints are not as strong. 

Conventional baryon-antibaryon annihilation typically produces a large number of 
pions which eventually decay though an intermediate muon and thus generate a 
significant number of neutrinos and antineutrinos in the range where 
SuperK has a high sensitivity. 
We will argue that the critical difference in the case of nuclear annihilations 
involving an antiquark nugget is that the annihilation proceeds within the 
colour superconducting (CS) phase where the energetics are significantly different. 
The main point is that in most CS phases the lightest pseudo Goldstone mesons 
(the pions and Kaons) have masses in the  20 MeV range \cite{Alford:2007xm,Rajagopal:2000wf} 
in huge contrast with the hadronic confined phase where $m_{\pi}\sim 140$ MeV.   
Therefore, the decay of light pseudo Goldstone mesons of the CS phase cannot produce  
neutrinos in the 20-50 MeV energy range and are not subject to the 
SuperK constraints employed in \cite{Gorham:2015rfa}. Some fraction of annihilations 
will occur sufficiently near the quark surface that the secondary mesons may form 
outside of the colour superconductor. As they produce the only potentially relativistic 
charged particles able to escape the nugget these surface events were important 
in assessing the ability of a nugget to produce an extensive air shower as in 
\cite{Lawson:2010uz}. However these event will represent only 
a small fraction of all annihilations and thus of total neutrino production. This fraction was 
not expressly calculated in \cite{Lawson:2010uz} which was primarily interested in 
the general details of shower geometry. We will argue here 
that in the physically well motivated scenario where only a few percent of annihilations 
result in this sort of near surface muon production the constraints imposed by 
SuperK are evaded.

\section{Quark nugget dark matter}
\label{sec:QNDM}
Before turning to the details of neutrino production we will first give a 
brief overview of the quark nugget dark matter model. 
The idea that the dark matter may take the form of composite objects of 
standard model quarks in a novel phase goes back to stranglet models \cite{Witten:1984}. 
In these models the presence of strange quarks stabilize quark matter at sufficiently 
high densities allowing strangelets formed in the early universe to remain stable 
over cosmological timescales. The quark nugget model is conceptually similar, with the 
nuggets being composed of a high density colour superconducting phase.
The only new crucial element proposed in \cite{Zhitnitsky:2002qa,Oaknin:2003uv}, 
in comparison with the previous studies \cite{Witten:1984} is that the nuggets could be 
made of matter as well as antimatter in this framework.  For further details see the original 
papers \cite{Zhitnitsky:2002qa, Oaknin:2003uv, Zhitnitsky:2006vt} 
as well as the recent short review \cite{Lawson:2013bya}. 

The original motivation for this model was as follows: 
It is generally assumed that the universe 
began in a symmetric state with zero global baryonic charge 
and later (through some baryon number violating process) 
evolved into a state with a net positive baryon number. As an 
alternative to this scenario we advocate a model in which 
``baryogenesis'' is actually a charge separation process 
in which the global baryon number of the universe remains 
zero. In this model the unobserved antibaryons come to comprise 
the dark matter in the form of dense nuggets of quarks and antiquarks.  
In this model baryogenesis occurs at 
the QCD phase transition. Both quarks and antiquarks are 
thermally abundant in the primordial plasma but, in 
addition to forming conventional baryons, some fraction 
of them are bound into heavy nuggets of quark matter in a 
colour superconducting phase. The formation of the  nuggets made of 
matter and antimatter occurs through the dynamics of collapsing axion domain walls.

If the fundamental $\theta$ parameter were identically zero at the early universe
QCD phase transition, equal numbers of nuggets 
made of matter and antimatter would be formed.  This would result in the vanishing 
of the visible baryon density at the present epoch and no baryogenesis effect. 
However, the fundamental CP violating processes associated 
with the $\theta$ term in QCD (which is assumed to be non-zero at the 
QCD phase transition) result in the preferential formation of 
antinuggets over nuggets\footnote{This preference  is essentially 
determined by the sign of $\theta$.}.  This source of strong CP violation is no longer 
available at the present epoch as a result of the dynamics of the axion, 
see e.g. recent reviews 
\cite{Asztalos:2006kz,Sikivie:2009fv,Rosenberg:2015kxa,Aprile:2014eoa} on this subject. 

The remaining antibaryons in the early universe plasma then 
annihilate away leaving only the baryons whose antimatter 
counterparts are bound in the excess of antiquark nuggets and are thus 
unavailable to annihilate. All asymmetry effects at $\theta \sim 1$ are order 
one which is precisely  why the resulting visible and dark matter 
densities must be the same order of magnitude 
\be
\label{Omega}
 \Omega_{\rm dark} \approx \Omega_{\rm visible}
\ee
as they are both proportional to the same fundamental $\Lambda_{\rm QCD} $ scale,  
and they both are originated at the same  QCD 
epoch\footnote{\label{insensitivity}The feature that this 
model produces visible and dark matter at roughly the same scale is 
largely insensitive to the absolute value
of the initial magnitude of the $\theta$ parameter. In other words, this model does 
not require any fine tuning of the initial  $\theta$ value.}.
In particular, the observed 
matter to dark matter ratio $\Omega_{\rm dark} \simeq 5\cdot \Omega_{\rm visible}$ 
corresponds to a scenario in which the number of antinuggets is larger than the 
number of nuggets by a factor of $\sim$ 3/2 at the end of nugget formation. This would 
result in a matter content with baryons, quark nuggets 
and antiquark nuggets in an approximate ratio 
\be
\label{ratio}
|B_{\rm visible}|: |B_{\rm nuggets}|: |B_{\rm antinuggets}|\simeq 1:2:3, 
\ee
with no net baryonic charge. If these processes 
are not fundamentally related the two components 
$\Omega_{\rm dark}$ and $\Omega_{\rm visible}$  could easily 
exist at vastly different scales. 
 
We present below a short overview describing 
a possible formation mechanism for these nuggets based 
on axion domain wall dynamics following \cite{Zhitnitsky:2002qa}, see also another 
proposal \cite{Atreya:2014sca} when formation of the nuggets is associated with 
drastically different types of the domain walls.  
 
Though the QCD phase diagram at $\theta\neq 0$ as a function of temperature and 
chemical potential  is basically unknown, it is well understood that $\theta$ is in fact an 
angular variable, and therefore supports various types of domain walls, including the 
so-called $N=1$ domain walls across which $\theta$ interpolates between one and 
the same physical vacuum state $\theta\rightarrow\theta+2\pi$. Furthermore, it is expected 
that closed bubbles made of these $N=1$ axion domain walls are also produced 
during the QCD phase transition
with a typical wall tension $\sigma_a\sim m_{a}^{-1}$ where $m_a $ is the axion mass. 
The collapse of these closed bubbles is halted due to the fermi pressure inside 
the bubbles which counters the wall tension.  The crucial element which stops the 
collapse of the bubbles from complete annihilation is the presence of the QCD substructure 
inside the axion domain wall.  This substructure forms immediately after the QCD phase 
transition as discussed in \cite{Zhitnitsky:2002qa}.  
The equilibrium of the obtained system has been analyzed in \cite{Zhitnitsky:2002qa} 
for a specific axion domain wall  tension 
within the observationally allowed window $10^{-6} {\rm eV}\leq m_a\leq  10^{-3}{\rm eV}$ 
consistent with the recent constraints  
\cite{Asztalos:2006kz,Sikivie:2009fv,Rosenberg:2015kxa,Aprile:2014eoa}.
It has also been argued in \cite{Zhitnitsky:2002qa}  that equilibrium is typically  achieved 
when the Fermi pressure inside the nuggets falls into the region in which colour superconductivity 
sets in\footnote{In this framework bubble formation does not require a first order 
phase transition as was required in the original strangelet proposal \cite{Witten:1984}. 
This is because the $N=1$ axion  domain walls form irrespective of the order of the phase 
transition.  The phase diagram in general and the order of the 
phase transition in particular at $\theta\neq 0$ are still unknown because of the well 
known ``sign problem" in the QCD lattice simulations at $\theta\neq 0$.}. 

The fact that the CS phase may be realized in nature in the cores of neutron stars has been 
known for some time \cite{Alford:2007xm,Rajagopal:2000wf}. A new element 
advocated in \cite{Zhitnitsky:2002qa}  is that a similar dense environment may be realized 
in nature if the axion domain wall pressure plays a role of a ``squeezer", 
similar to the gravitational pressure in the neutron star physics. 

Using dimensional arguments one can easily infer that the size and the baryon 
charge of the nuggets must scale with the axion mass as follows
\be
\label{sigma}
\sigma_a\sim m_a^{-1}, ~~~R\sim \sigma_a, ~~~~B\sim \sigma_a^3.
\ee
Therefore, as the axion mass $m_a$ varies within the observationally allowed window 
$10^{-6} {\rm eV}\leq m_a\leq  10^{-3}{\rm eV}$ the nuggets parameters  
vary as follows
\be
\label{B}
10^{-6} {\rm cm}\lesssim R \lesssim 10^{-3}{\rm cm}, ~~~ 10^{23}\lesssim B\lesssim 10^{32}.
\ee
This essentially determines the allowed parameter space for the quark nuggets. Much of 
the corresponding allowed region is essentially unconstrained by present  
experiments, as will be discussed below. 

Another fundamental ratio (along with 
$\Omega_{\rm dark} \simeq 5 \Omega_{\rm visible}$  discussed above)
is the baryon to entropy ratio at present time
\be
\label{eta}
\eta\equiv\frac{n_B-n_{\bar{B}}}{n_{\gamma}}\simeq \frac{n_B}{n_{\gamma}}\sim 10^{-10}.
\ee
In our proposal (in contrast with conventional baryogenesis frameworks) this ratio 
is determined by the formation temperature $T_{\rm form}$ at which the nuggets and 
antinuggets compete their formation and below which annihilation does 
not significantly affect the nuggets' composition. As the present day baryon to 
photon ratio (\ref{eta}) is very small this model is not highly sensitive to  the efficiency 
of nugget formation provided it is still larger than $10^{-10}$. 
Following their initial formation the nuggets remain coupled 
to the baryon plasma until it has cooled to $T_{\rm form}$ and it is this decoupling 
that fixes the ratio (\ref{eta}) rather than the original distribution of nuggets, 
as specified in footnote (\ref{insensitivity}) this result is also insensitive to the initial 
value of the $\theta$ parameter.
This temperature is determined by many factors: transmission/reflection coefficients, 
evolution of the nuggets, expansion of the universe, cooling rates, evaporation rates, the  
dynamics of the axion domain wall network, etc. All these effects, in general, may equally 
contribute to determining $T_{\rm form}$ at the QCD scale. Technically, the corresponding 
effects are hard to compute from the first principles as even the basic properties of the  
QCD phase diagram at $\theta\neq 0$ are still unknown. However, the estimate 
of $T_{\rm form}$ within a factor of 2 is quite a simple exercise as  $T_{\rm form}$  
must be proportional to the gap $\Delta\sim 100$ MeV when the CS phase sets in  
inside the nuggets.  The observed ratio (\ref{eta}) corresponds to 
$T_{\rm form}\simeq 41$ MeV, see \cite{Oaknin:2003uv} for the  details.
This temperature indeed represents a typical QCD scale,  slightly below 
the critical temperature $T_c\simeq 0.6 \Delta\simeq 60$ MeV,   according to  
standard  estimates  for colour superconductivity, see  reviews  
\cite{Alford:2007xm,Rajagopal:2000wf}. 

Unlike conventional dark matter candidates, such as WIMPs 
(Weakly interacting Massive Particles) the dark-matter/antimatter
nuggets are strongly interacting and macroscopically large.  
However, they do not contradict any of the many known observational
constraints on dark matter or
antimatter  for three main reasons~\cite{Zhitnitsky:2006vt}:
\begin{itemize} 
\item They carry a huge (anti)baryon charge 
$|B|  \gtrsim 10^{25}$, and so have an extremely tiny number
density; 
\item The nuggets have nuclear densities, so their effective interaction
is small $\sigma/M \sim 10^{-10}$ ~cm$^2$/g,  well below the typical astrophysical
and cosmological limits which are on the order of 
$\sigma/M<1$~cm$^2$/g;
\item They have a large binding energy such that the baryon charge locked in the
nuggets is not available to participate in big bang nucleosynthesis
(\textsc{bbn}) at $T \approx 1$~MeV. 
\end{itemize} 

There are currently a number of both ground based and astrophysical 
observations which impose constraints on  allowed quark nugget dark matter 
parameters. These include the non-detection of a nugget flux by the IceCube 
monopole search \cite{Aartsen:2014awd} which limits the flux of nuggets to 
$\Phi_N < 1$km$^{-2}$ yr$^{-1}$. Similar limits are likely also obtainable 
from the results of the Antarctic Impulse Transient Antenna (\textsc{ANITA}) 
\cite{Gorham:2012an}.
It has also been estimated in \cite{Abers:2007ji} that, based on Apollo data, 
nuggets of mass from $\sim$ 10 kg to 1 ton (corresponding to 
$B \sim 10^{28\text{-}30}$) must account for less than an order of
magnitude of the local dark matter.  While our preferred range of
$B\sim 10^{25}$, see below,  is somewhat smaller and is not excluded 
by \cite{Abers:2007ji}, we still believe that $B\geq 10^{28}$ is not
completely excluded by the Apollo data, as the corresponding
constraints are based on specific model dependent assumptions about
the nugget mass-distribution. 

While ground based direct searches   
offer the most unambiguous channel for the detection of quark nuggets 
the flux of nuggets is inversely proportional to the nugget mass and 
consequently even the largest available detectors are incapable of 
excluding a nugget flux across their entire potential mass range. 

To reiterate: the weakness of the visible-dark matter interaction is achieved 
in this model due to the small geometrical parameter $\sigma/M \sim B^{-1/3}$ 
rather than due to a weak coupling of a new fundamental field to standard model particles. 
In other words, this small effective interaction $\sim \sigma/M \sim B^{-1/3}$ 
replaces a conventional requirement
of sufficiently weak interactions of the visible matter with WIMPs. 

The mean baryon number of the nuggets $\langle B\rangle$ is, in principle, 
calculable from first principles. However, such a calculation depends on the 
strongly coupled, non-equilibrium physics of the QCD phase transition and is 
subject to large uncertainties. As such, we treat $\langle B\rangle$ as a parameter 
to be constrained observationally with larger $\langle B\rangle$ values 
producing weaker observational consequences as discussed above. 
Any such consequences will be largest where the densities 
of both visible and dark matter are largest such as in the 
core of the galaxy or the early universe. In other words, the nuggets behave like  
conventional cold dark matter in environments where the density of the visible matter is small, 
while they become interacting and radiating objects (i.e. effectively become visible matter)  
if they enter an environment of sufficiently large density.

A variety of astrophysical observations suggest that, in order to avoid over producing  
the observed galactic diffuse background the nuggets must have a sufficiently large 
$\langle B\rangle > 10^{24}$.  It should be noted that the galactic spectrum 
contains several excesses of diffuse emission the origin of which is unknown, the best 
known example being the strong galactic 511~keV line. If the nuggets have a baryon 
number in the $\langle B\rangle \sim 10^{25}$ range they could offer a 
potential explanation for several of 
these diffuse components. For further details see the original works 
\cite{Oaknin:2004mn, Zhitnitsky:2006tu,Forbes:2006ba, Lawson:2007kp,
Forbes:2008uf,Forbes:2009wg,Lawson:2012zu,Lawson:2015xsq}.
In all these cases photon emission originates 
from the outer layer of the nuggets known as the electrosphere. 
The properties of the electrosphere 
are not very sensitive to the specific features of the CS phase realized in the dense 
nugget core. By contrast, in our present study the structure of the nugget's core plays 
the crucial role as neutrino emission originates from within the dense CS matter.  
We review the features of the CS phase relevant to neutrino emission in the next section.

\section{Colour Superconductivity and Quark Nuggets}\label{cfl}
There are many possible CS phases due to the generation of a gap $\Delta$ through 
different channels with slightly different properties.
While the relevant physics is a part of the standard model, QCD with no free parameters, 
the corresponding phase diagram is still a matter of debate as it strongly depends on 
the precise numerical value of the gap $\Delta$, see review articles 
\cite{Alford:2007xm,Rajagopal:2000wf}.
For our purposes though the key characteristics are very much the same for all
phases. Therefore, we limit ourself below to reviewing the most developed, 
the so called CFL (colour flavour locking) phase.  

\subsection{Quasi-particles in the CFL phase}\label{quasiparticles}
As already mentioned, the physics of photon radiation is dominated by the 
low-energy degrees of freedom present at the quark matter surface and in the electrosphere.  
The neutrino emission is entirely determined by the dense quark matter realized in the nugget's 
core. Within the nuggets, the picture is of some charged phase of
colour-superconducting matter with a gap of some $\Delta\simeq 100$ MeV, and with a
free Fermi-gas of positrons to maintain neutrality. In order to discuss neutrino production  
in this context we will first enumerate the basic excitations of the CS phase the decay 
of which will give rise to the neutrino spectrum. 
\begin{description}
\item[Quarks and Gluons] The fundamental excitations in quark matter
at sufficiently high densities are the quarks 
(similar to baryon octet $ p, \Sigma, \Xi $ of the conventional hadronic phase) 
and gluons (similar to vector meson octet $\rho, \omega, \phi, K^*$ of the hadronic phase).  
In contrast with the hadronic phase, however, these
excitations are coloured. They are strongly interacting quasi-particles and have ``masses'' 
on the order of the superconducting gap $\sim 100$ MeV which is  much smaller than the typical 
mass scale $\sim $ 1 GeV of the excitations (the hadrons) in the confined phase.
Although the initial annihilation will produce these quasi-particles, the
energy will ultimately be transferred to lower energy degrees of freedom which are the 
Nambu-Goldstone Bosons.
 
As they act as only a brief intermediate phase of the decay chain, we shall not 
elaborate on detailed properties of these relatively heavy strongly 
interacting quasi-particles. These properties are not essential in determining the neutrino 
production spectrum, which is the main subject of this work. The only comment we 
would like to make here is that the splitting between different quark states 
(similar to $ p, \Sigma, \Xi $) is of order $m_s^2/(2\mu)\sim 15 $ MeV where $m_s$ 
is the strange quark mass and $\mu$ is the quark chemical potential, see  
for example \cite{Alford:2007xm}.  This represents a tiny energy in comparison with the 
splitting in the conventional hadronic phase where the relevant scale is determined by 
the strange quark mass $m_s$ itself and is of order 150 MeV. 
These estimates suggest that even in rare events when a heavy quark decays 
to a lighter cousin with emission of a neutrino, the corresponding neutrino's  energy will 
not exceed 10 MeV  in the CFL phase. 

\item[Pseudo--Nambu-Goldstone Bosons (Mesons)] The spontaneous
breaking of chiral symmetry in colour-superconductors gives rise to
low-energy pseudo--Nambu-Goldstone modes with similar quantum
numbers to the mesons (pions, kaons, etc.).  These objects, however, are
collective excitations of the CS state rather
than vacuum excitations as is the case for the conventional confined hadronic phase.
The finite quark masses explicitly break chiral symmetry, giving rise to these
``pseudo''--Nambu-Goldstone modes on the order of 10 -20 MeV, in huge contrast 
with the hadronic confined phase where the lightest mass meson has 
$m_{\pi}\simeq 140$ MeV. However, the hierarchy between the lightest 
Nambu-Goldstone (NG) and vector meson masses approximately holds 
for both: CFL and hadronic phases with  $m_{NG}/m_{\rho}\approx 1/5$.
  
To be more precise, we consider large $\mu$ limit for which the masses and 
other relevant parameters in the CFL phase can be explicitly computed 
\cite{Alford:2007xm,Rajagopal:2000wf}: 
\be
\label{NG}
m^2_{\pi^{\pm}}&\simeq&\frac{2c}{f_{\pi}^2}m_s(m_u+m_d), ~~ f_{\pi}^2\sim \mu^2\nonumber\\
m^2_{K^{\pm}}&\simeq&\frac{2c}{f_{\pi}^2}m_d(m_u+m_s) , ~~~ c\simeq \frac{3\Delta^2}{2\pi^2} \nonumber\\
m^2_{K^0}&\simeq&\frac{2c}{f_{\pi}^2}m_u(m_d+m_s), ~~~ v_{NG}^2=\frac{1}{3}.
\ee
In these formula $f_{\pi}\sim 100 $ MeV is proportional to the chemical potential 
and plays the same role as the pion decay constant ($f_{\pi}$) in the conventional 
confined hadronic phase;  $v_{NG}$ deviates from speed of light due to the 
explicit violation of the Lorentz invariance in the system
such that the dispersion relations for all quasiparticles are drastically different 
from their counterparts in vacuum. 
The modes in (\ref{NG}) are all strongly interacting and include both charged
($K^{\pm}$, $K$, $\bar{K}$, $\pi^{\pm}$) and truly neutral ($\pi^{0}$,
$\eta$, $\eta'$) modes. These neutral modes get mixed but, while the corresponding 
mass matrix is also known, our discusssion does not require an explicit expression for 
this matrix, as the typical scale remains the same as presented in (\ref{NG}).   
  
There are a few very important differences (aside from $v_{NG}\neq 1$) between 
the NG modes of the CFL phase and those of the confined hadronic phase. 
First of all, the NG bosons are much lighter than in vacuum. This is because their 
masses are proportional to $m_q^2$ rather than to $m_q$, as at zero chemical potential. 
In addition, there is a further suppression by a factor of $\Delta/\mu \ll 1$ which arises 
because the NG bosons are collective excitations of the diquark condensate, in 
contrast with the chiral vacuum condensate of the confined phase. As a result the kaon 
of the CS phase is lighter than the CS pion, by a factor of $m_d/(m_d+m_u)$. 
As a result, the lightest NG meson, the kaon, has a mass in the range of 5 to 20 MeV 
depending on $\Delta$ and taking $\mu\simeq 400$ MeV, see 
for example \cite{Rajagopal:2000wf}.  
  
Another important difference between the NG modes in dense matter and in vacuum
is in the dispersion relations for the NG modes which assume the following form, 
see for example \cite{Alford:2007xm}:
\be
\label{NG1}
E_{K^{\pm}}&=&\mp\frac{m_s^2}{2\mu}+\sqrt{v_{NG}^2 p^2+m^2_{K^{\pm}}} \nonumber\\
E_{K^{0}}&=&-\frac{m_s^2}{2\mu}+\sqrt{v_{NG}^2 p^2+m^2_{K^{0}}} \nonumber\\
E_{\bar{K}^{0}}&=&+\frac{m_s^2}{2\mu}+\sqrt{v_{NG}^2 p^2+m^2_{K^{0}}} \nonumber\\
E_{\pi^{\pm}}&=& \sqrt{v_{NG}^2 p^2+m^2_{\pi^{\pm}}}, 
\ee
such that the rest energy of the lightest NG mesons does not exceed 
the 10-20  MeV range. In fact $E_{{K}^{0}}$ may even vanish, in which 
case the ${K}^{0}$ field forms a condensate (the so-called $CFL~ K^0$-phase). 
We shall not elaborate on this topic in the present paper but instead refer the interested 
reader to the review  article  \cite{Alford:2007xm} and the large number of  
references on original works cited there.  
  
One should comment here that the dispersion relations for the NG modes within the  
anti-nuggets can be obtained from  (\ref{NG1}) by replacing  $\mu\rightarrow  -\mu$ 
such that the lightest NG states become $\bar{K}^{0}$ and $K^-$ for  
nuggets made of antimatter. 
  
The drastic differences in properties between the NG modes in dense matter and in 
vacuum reviewed above play the key role in our discussions of the neutrino emission 
from CS nuggets, which is the subject of  section \ref{neutrino}.

\item[Nambu-Goldstone Phonon] There is one truly massless
Nambu-Goldstone mode which is associated with spontaneous baryon number
violation.  This mode, however, carries neither colour nor electric
charge.  Thus, it is extremely weakly coupled to the other NG modes and plays an
insignificant role in the  processes we consider here. Nevertheless, it may also emit 
neutrinos. Therefore,  it will also be briefly discussed in section \ref{neutrino} along 
with the main neutrino emitters, the lightest NG states.

\end{description}

\subsection{Energy Transfer}\label{sec:energy-transfer}
Now we consider how energy is transferred from nucleon annihilation
events to radiation and neutrino emission.  Nucleons impinging upon antimatter nuggets
will penetrate some characteristic depth $d\ll R$ much less than the
radius of the nuggets.  The depth will be set by the nuclear
interaction scale: For example, if the annihilation
$\text{p}^{+}\text{p}^{-}$ cross-section is taken to be the vacuum
value, the proton will have a lifetime of only 2~fm/$c$.  However, as argued in 
\cite{Forbes:2006ba}  the lifetime
could be considerably larger in the CS phases due to the 
coherence requirement for annihilation,  whereby the quantum numbers
for all three antiquarks must be correlated for a successful
annihilation with the incoming proton, in this case the lifetime could be 
as large as 100~fm$/c$. 

A simple way to explain the argument is to consider QCD with a large number of colours $N_c$.
In this case a colour singlet incoming nucleon entering the CS phase must pick up  
three antiquarks for successful annihilation. These three antiquarks cannot have 
arbitrary  colour properties, but rather must be precisely organized in 
an appropriate colour singlet combination. While the choice for the first antiquark is 
arbitrary, the choice for an appropriate colour for the second and third quark
would lead to the suppression factor $N_c^{-1}\cdot N_c^{-1}$. 
This suppression effect is similar in many respects 
to a phenomenon which is known to occur for (composite) soliton- antisoliton 
annihilation   in condensed matter physics. 
Generically, the probability for the soliton- antisoliton elastic scattering is much larger 
than the probability for annihilation\footnote{Similarly, the composite magnetic 
monopole- antimonopole  annihilation rate is much smaller than a typical dimensional 
analysis (represented by the size of the system) would suggest, 
see e.g.\cite{Vachaspati:2015ahr} and references therein.}.   
Furthermore, the momentum distribution for quarks from a nucleon in vacuum 
is drastically different from momentum distribution in the CS phase where momentum 
counts from the fermi surface. The corresponding momentum wave function overlap 
will thus also be suppressed. 
In fact, the tendency of increasing lifetime with increasing chemical potential has been 
previously discussed  for relatively low densities in  the case of annihilation of anti-nucleons 
in nuclear matter \cite{Mishustin:2004xa}. In that paper it has  been argued that the 
lifetime could be considerably larger (possibly even 10-30 fm/c) than the fm/c scale 
observed in vacuum. 

In any case, on the scale of the nuggets, the annihilations will take place
within $d\sim 100$~fm or so from the surface, which is still within
the regime where the quark matter is charged.  Thus, the annihilations
take place in a region of quark matter where there are positrons 
present to maintain beta-equilibrium and neutrality.  These positrons
are the lightest modes that couple strongly and will thus ultimately
carry most of the annihilation energy as argued in  \cite{Forbes:2006ba}.

The nucleon annihilation will initially produce a handful of high
energy strongly interacting gluons and quarks with properties as described in 
the previous section \ref{quasiparticles}.  These will stream away from the annihilation site
with about half streaming towards the surface of the nuggets,
and ultimately transfering their energy to high-energy positrons.
This portion of the energy released as a result of annihilation is emitted as 
electromagnetic radiation in the form of diffused x rays
as discussed in \cite{Forbes:2006ba}, and shall not be discussed further in this work. 
The other half, however, will proceed deep within the nuggets.
 
The  strongly interacting gluons and quarks which move 
towards the nugget's core will rapidly decay into pseudo--Nambu-Goldstone 
collective modes (similar to $\rho\rightarrow 2\pi $ decays in hadronic matter), 
distributing energy down to their mass scale of $20 $ MeV or so.    
Thus, through this mechanism, some fraction  of the
initial  2 GeV  annihilation energy will ultimately be transferred to the NG modes.
These NG modes will then undergo weak decays emitting the neutrinos 
which are the focus of the present work.

The goal of the next section is to advocate that the energy of  the majority 
of these neutrinos will  
not exceed 20 MeV. This claim is in drastic contrast with
hadronic phase where the $\pi$ and $K $ mesons generically emit highly energetic 
neutrinos in the 50 MeV range through direct production or through $\mu$-lepton  
emission and subsequent decay into energetic neutrinos. 

\subsection{Neutrino emission from  NG bosons}\label{neutrino}
The neutrino emission from CFL phase quark matter has been studied previously 
in a number of papers mostly in context of the physics of neutron stars, see the 
original papers \cite{Jaikumar:2002vg,Reddy:2002xc,Reddy:2003ap} and the review 
article \cite{Alford:2007xm}. In this subsection we will elaborate on estimates of 
neutrino emission from the superconducting quark nuggets and anti-nuggets. 

First, we want to argue that the heaviest NG mesons (the $K^-$ and $\bar{K^0}$ from 
(\ref{NG1})), which might be as massive as 80 MeV according to \cite{Reddy:2002xc},  
and which may potentially produce sufficiently energetic neutrinos to violate the SuperK 
constraints at energies $> 20 $ MeV, generically do not produce such energetic neutrinos. 

There are a few suppression factors which lead to this conclusion. First of all, 
the production of strange $K$ mesons is suppressed due to the so-called ``Zweig rule" 
when a $\bar{s}s$ pair must be produced from the ground state as a result of annihilation. 
This suppression factor must be proportional to $1/N_c^2\sim 0.1$ where $N_c=3$ is the 
number of colours in QCD. This suppression factor is well understood theoretically and well 
established experimentally in numerous $p\bar{p}$ and $\bar{p} A$  experiments at low and 
high energies when the mass difference between strange and non-strange particles   
does not play a role. We expect this suppression to hold even when $K$ mesons 
are lighter that the $\pi$ mesons.  

The next  argument behind this claim is the observation that the heaviest NG mesons  
mainly decay to their lighter NG cousins 
(similar to $K\rightarrow \pi\pi$ in hadronic phase)  with a much larger probability 
than to an electron accompanied by neutrinos, $K^{\pm}\rightarrow e^{\pm} \nu$. 
The basic reason is that the corresponding matrix elements are always suppressed by  
the electron mass\footnote{\label{suppression}This statement is correct for temperatures 
$T\simeq 0$ close to zero. For non vanishing temperature there is an extra contribution 
which does not have a vacuum counterpart. This contribution is generated by 
the explicit violation of Lorentz invariance in the system when the NG fields couple 
to the temporal and the spatial components of the axial vector current with different constants. 
However, this extra contribution does not modify our claim because the typical 
temperature of the nuggets 
considered in the present work is small, $T\ll m_e$, see section \ref{T} for the estimates.}.  As a 
result, the heaviest NG modes in the CFL phase to not generically decay with the emission 
of energetic neutrinos. This is of course in huge contrast with the conventional pattern in 
the hadronic phase where leptonic decays  $K^{\pm}\rightarrow \mu^{\pm}\nu$  
represent the main channel of the $K^{\pm}$ decays. The point is that even the heaviest 
NG mesons in the CFL phase are still much lighter than the muon. Therefore, this $\mu^{\pm}\nu$  
channel of $K$ decay is kinematically forbidden. 

The 3-particle decays (similar to $K^{\pm}\rightarrow \pi^0 e^{\pm} \nu$ in hadronic phase) 
are not strictly kinematically forbidden, and not suppressed by a factor of $m_e$. Nevertheless, it 
is expected to be smaller (at least by a factor of 
$\sim 0.2$) in comparison with the dominant $K\rightarrow 2 \pi$ decay in 
analogy with a  similar 3-particle phase volume suppression in vacuum. Furthermore, 
a typical energy for the neutrino in such 3-particle decay is quite small as the largest chunk of 
the available energy goes to the pion mass which  is of order 
30 MeV according to \cite{Reddy:2002xc}. In other words, the peak of the 
intensity distribution is achieved at 
$\sim \frac{1}{3} (80-30) \rm{MeV}~\sim 16 ~\rm{MeV}$, 
while the probability of finding a neutrino with a larger energy ($> 20 $ MeV) is   suppressed.
This factor  is an addition to the suppression factors (due to the Zweig rule 
and 3-particle phase volume) already mentioned above. 

These generic arguments suggest that the majority of neutrinos will be emitted by the lightest  
rather than the heaviest NG bosons, similar to  $\pi^{\pm}\rightarrow e^{\pm} \nu$. 
In this case the energy of the emitted neutrinos is in range of 15 MeV for the CFL phase. 
In addition, as we argued above, muons cannot be produced at all 
in the CFL phase for purely kinematical reasons. Therefore, the energetic neutrinos which 
played  the key role in analysis of \cite{Gorham:2015rfa} and which are normally produced in 
the $\mu^{\pm}\rightarrow e^{\pm}\nu_e\nu_{\mu}$ decay channels and are not generally 
produced when these decays occur within the CS matter.  This it the essential point of the 
present analysis. 

For completeness of our analysis of neutrino emission we will also make few 
comments on the process $\pi^0\rightarrow \nu\bar{\nu}$ which is strongly suppressed  
in vacuum, but may occur in quark dense matter. The corresponding decay, in terms of 
kinematics, is similar to the $K^{\pm}\rightarrow e^{\pm} \nu$ decay mentioned above. 
Therefore, it must also be proportional to the lepton mass in vacuum, $m_{\nu}^2$  
which is very close to zero. This is precisely the source of the strong suppression of 
the $\pi^0\rightarrow \nu\bar{\nu}$  decay in vacuum. However, in the CFL phase the 
relevant matrix element is not proportional to $m_{\nu}$ due to the explicit violation of 
Lorentz invariance, similar to the extra term in the $K^{\pm}\rightarrow e^{\pm} \nu$ decay 
mentioned above (see footnote \ref{suppression} for a few comments and also the references 
\cite{Jaikumar:2002vg,Reddy:2002xc} where the corresponding computations were 
originally performed). Nevertheless, this type of  suppression still holds in our case  
due to relatively low temperature $T\ll m_e$ of the nuggets while travelling  
thorough the sun, as will be argued in the following section. 

Finally, neutrinos might be also produced in CFL phase through the process 
$\phi+ \phi\rightarrow \phi+\nu+\bar{\nu}$ where $\phi$ is the truly massless NG  
boson  associated with spontaneous baryon number violation.
The corresponding process is highly suppressed \cite{Jaikumar:2002vg} as the 
amplitude is proportional to the temperature
which is very small in this problem as estimated in the next section. 
  
\subsection{Temperature of nuggets travelling through the sun}\label{T} 
We follow ref. \cite{Forbes:2008uf} in our estimation of the temperature of the 
nuggets as they travel through the sun.
The basic idea is to equate the total surface emissivity $ F_{\text{tot}}$ with 
the rate at which annihilations deposit energy $F_{\text{ann}}$
within a given environment. The total surface
emissivity has been computed in \cite{Forbes:2008uf}. It is mostly determined 
by photon emission from the electrosphere, and it is given by
\begin{equation}
\label{eq:P_t}
F_{\text{tot}} =  \frac{\d{E}}{\d{t}\;\d{A}} \simeq  
\frac{16}{3} \frac{T^4\alpha^{5/2}}{\pi}\sqrt[4]{\frac{T}{m}}\\
\end{equation}
In order to maintain the overall energy balance, the nuggets must emit energy at
the same rate that it is deposited through proton annihilation,
\begin{equation}
\label{eq:FtotFann}
F_{\text{tot}} = (1-g)F_{\text{ann}}
= (1-g)\frac{\d{E_{\text{ann}}}}{\d{t}\,\d{A}},
\end{equation}
where $1-g$ is the fraction of the annihilation energy that is
thermalized (and thus $g$ is the fraction of energy emitted directly by excited 
positrons.)  Note that both the rate of emission and the rate of
annihilation are expressed per unit surface area, so that the equilibrium
condition is independent of the nuggets' size, and therefore of their average 
baryon number $\langle B\rangle $. 

The rate of annihilation $F_{\text{ann}}$ is
\begin{equation}
\label{eq:Fann1}
F_{\text{ann}} = 2\GeV\cdot f\cdot v \cdot n_{\rm sun} 
\end{equation}
where $2\GeV = 2\,m_{p}$ is the energy liberated by proton annihilation,
$v$ is the average speed  of the nugget through the sun,
$n_{\rm sun}$ is the average nucleon density in the sun, and $f\sim 0.1$
is a suppression  factor  due to the possibility of reflection from
the sharp quark-matter surface.

The typical galactic scale for the speed is $v\sim 300 \text{ km/s}
\sim 10^{-3} c$, while the density  $n_{\rm sun}\sim 10^{24} {\rm cm}^{-3}$

Combining these numerical values, we obtain
\begin{equation}
\label{eq:Fann2}
F_{\text{ann}} \sim \frac{10^{30}~\text{GeV}}{\text{cm}^2\cdot\text{s}}
\cdot \left(\frac{f}{10^{-1}}\right) \cdot \left(\frac{v}{10^{-3}c}\right)\cdot
\left(\frac{n_{\rm sun}}{10^{24}/\text{cm}^{3}}\right)
\end{equation}
which must be compared with the total surface emissivity~(\ref{eq:P_t}) 
which in appropriate units can be represented as 
\begin{equation}
\label{eq:Ftot}
F_{\text{tot}} \sim 10^{30}
\frac{\text{GeV}}{\text{cm}^2\cdot\text{s}}
\left(\frac{T}{\text{
0.1 MeV}}\right)^{4+1/4}.
\end{equation}
Taking the typical values for these parameters mentioned above 
and comparing (\ref{eq:Fann2}) with (\ref{eq:Ftot}) we arrive at an estimate of   
a temperature of $T\simeq 0.1 \MeV$ for anti-nuggets travelling through the sun. 
 
This numerical value for the temperature $T\simeq 0.1 \MeV$  is important in making a few 
conclusions. First, the relatively heavy NG bosons with rest energy  $\sim 80 \MeV$ do not 
generically decay to energetic neutrinos $K\rightarrow e \nu$ as the conventional suppression 
factor $\sim m_e^2$ still holds even for this non-zero temperatures because the relevant 
temperature is quite low as we argued above, $T\ll m_e$.  Second, the estimate 
$T\simeq 0.1 \MeV$ represents a typical width for 2-particle decays such as 
$\pi \rightarrow e \nu$ or $\pi^0 \rightarrow \nu\bar{\nu}$, at zero temperature this 
width would be practically zero as the energy is unambiguously 
fixed by the mass of the decaying particle. 
 
We end  this section with the following important conclusion: most of the neutrinos 
produced in the CFL phase are a result of the decay of the lightest $\pi^{\pm}$ mesons. Therefore, 
the typical energy of the emitted neutrinos is expected to be in  the 10-15 MeV range, in 
contrast with the 50 MeV scale associated with conventional hadronic decays as 
used in the analysis of \cite{Gorham:2015rfa}. 

Qualitatively our argument may be understood as follows:  the typical mass scale of 
the NG bosons in the CFL phase is 5-10 times smaller than in vacuum. Therefore,  we expect 
that the average number of NG bosons produced per annihilation event 
(releasing $\sim 2 $ GeV energy per event) is a factor of 5-10 
larger than in vacuum. Thus, the average number of emitted neutrinos per annihilation 
event will also be a factor of 5-10 larger than in vacuum.  The average energy of these 
neutrinos is set by the NG mass and is thus smaller by the same factor of 5-10. This difference 
in the properties of the neutrino spectrum plays the key role in any assessment of the 
predictions of the neutrino flux from antiquark nuggets in terms of constraints on the 
observed neutrino flux. 
 
\section{Neutrino flux constraints} \label{sec:Flux}
The capture of quark nuggets by the sun and their subsequent annihilation 
will produce an additional component in the solar neutrino spectrum. For 
comparative purposes we will adopt the solar capture rate used in \cite{Gorham:2015rfa} 
which was predicted to generate an antineutrino flux on the order of 
$\sim 10$ cm$^{-2}$s$^{-1}$. As argued  above this number could easily be a factor 
of ten larger due to the larger number of light NG states produced. As such, we will   
consider the possibility that the total flux of neutrinos generated by the annihilation of 
quark nuggets within the sun could be as large as, 
\begin{equation}
\label{eq:neutflux}
\Phi_{\rm tot} \approx 100 ~{\rm{cm}}^{-2} {\rm{s}}^{-1} .
\end{equation}
Of these roughly half will be neutrinos and half will be antineutrinos. As established 
in section \ref{neutrino} the majority of these will be produced by the decay of 
light NG bosons and carry an energy of $\sim M_{NG}/2$ and a decay width of 
$\Delta \sim T \sim 0.1$MeV. These decay lines will be accompanied by a 
much smaller continuum emission from three body decays. There is also a contribution 
due to the small probability of annihilation of protons directly on the interface 
between the vacuum and the  CS nuggets which may produce 
an energetic muon. For now this process will be assumed to make a negligible 
contribution to the total neutrino production rate and be lost in the three body continuum, 
but see our Conclusion for some further discussion of this process. 

\subsection{Neutrino emission model}\label{sec:nu_model}

While the exact structure of the various excitations of the CS is not known we may 
use expressions \ref{NG} and \ref{NG1} to draw some general conclusions. A specific choice 
of $\mu$ and $\Delta$ determines the mass scales for the light NG modes. As argued 
above these parameters are generally taken to be $\mu \sim 400$ MeV and 
$\Delta \sim 100$ MeV in the CFL phase.The splitting between the $K^+$ and $K^-$ 
states is then $m_s^2/\mu$. In an antiquark CS the light $K^-$ will 
produce a decay line in the antineutrino spectrum as will the heavier $\pi^-$ as its 
three body decay to a final state including a $K^-$ is Zweig suppressed even if it is 
energetically allowed. Zweig suppression also results in the production of a larger 
number of pions rather than of kaons such that the $\pi^-$ decay line will be stronger 
than the $K^-$ decay line by roughly a factor of ten and will contain the largest fraction 
of antineutrino produced. This line will be narrow, with thermal broadening at the level 
of $\sim 0.1$MeV. In order to compare to experimental results we will introduce a 
further broadening at the 1MeV scale corresponding to typical detector energy resolution. 
The antineutrino spectrum will also include continuum emission 
from the weak decay of heavier charged states. Though, as argued above this process 
is considerably disfavoured relative to decays to the lighter NG states. For demonstrative 
purposes we will assume that these processes account for one percent of antineutrino 
production, though the actual fraction could easily be smaller than this value. 
Larger chemical potentials correspond to lighter NG modes while the masses increase 
with the size of the gap $\Delta$. As neutrino flux constraints generally become more 
stringent with increasing energy we consider the case $\mu = 350$MeV and 
$\Delta = 150$MeV which corresponds to relatively heavy $K^-$ and $pi^-$ modes and 
thus imposes more conservative constraints. 
The antineutrino spectrum resulting from this choice of parameters as well as the 
relevant constraints (to be discussed further below) are shown in figure 
\ref{fig:antinu_spec}. 

The neutrino spectrum is slightly different from that of the antineutrinos due to the 
$K^+ / K^-$ mass splitting. The relatively heavy $K^+$ will tend to decay to 
a $\pi^+$ rather than directly producing a neutrino. Thus, in the neutrino spectrum 
we predict a single line from $\pi^+$ decays. As before this line will has an intrinsic 
width at the $\sim 0.1$MeV level, but will be broadened to match detector resolution. 
As with the antineutrino spectrum we will also assume the presence of a weaker 
continuum at the 1\% level. The neutrino spectrum, assuming the $\mu$ and 
$\Delta$ values discussed above is shown in figure \ref{fig:nu_spec}. 

\subsection{Observational constraints} 
\exclude{
For the purpose of our estimates we will assume a 
pair of gaussian decay lines (from the  NG bosons in CFL phase in anti-quark 
nuggets, the $K^{-}$ and $\pi^{\pm}$) with 
the $\pi^{\pm}$ channel favoured by as factor of $\sim10$ due to 
Zweig suppression of $K$ production. We will also introduce a broadening of these relatively 
sharp neutrino lines to match the $\sim 1$ MeV energy resolution available in the analysis 
of detector data. The continuum emission will taken to 
be suppressed by a factor of $\xi$ relative to the number of neutrinos 
in the decay lines from low momentum light NG 
decays. In principle the three particle decay spectrum will extend from very low 
energies up to $(M_{K} -M_{\pi})/2 \sim 30$ MeV while the even smaller direct muon 
spectrum (from muons produced on the surface) may extend as high as $\sim M_p$.} 
Relatively high energy neutrinos produced in the three body decays of heavy modes are  
subject to the antineutrino flux limits imposed by SuperK \cite{Lunardini:2008xd} 
and cited by \cite{Gorham:2015rfa}.
Specifically, the antineutrino flux limit of \cite{Lunardini:2008xd} 
is $\Phi_{\bar{\nu_e}} = 1.4 - 1.9$ cm$^{-2}$s$^{-1}$ for electron antineutrinos with 
energies above the 19.3 MeV threshold. This constraint is stronger than that 
imposed on electron neutrinos ($\Phi_{\nu_e} = 73.3 - 154$ cm$^{-2}$s$^{-1}$ in the 
same energy range) due to the relatively large inverse $\beta$-decay cross-section. 
Thus, while a roughly equal number of neutrinos and antineutrinos are produced 
the antineutrino constraints 
provide a stronger limit on nugget annihilation in this energy band. 
Based on this maximum flux we obtain an upper limit on the fraction of antineutrinos 
produced via three body decays with energies above the 19.3 MeV on the level of 
$\xi \leq 5 \%$. This value is certainly compatible with the known 
properties of most CS phases as it is suppressed by a number of  
factors as argued in  section \ref{neutrino}.

\begin{figure}
\includegraphics[width=\linewidth]{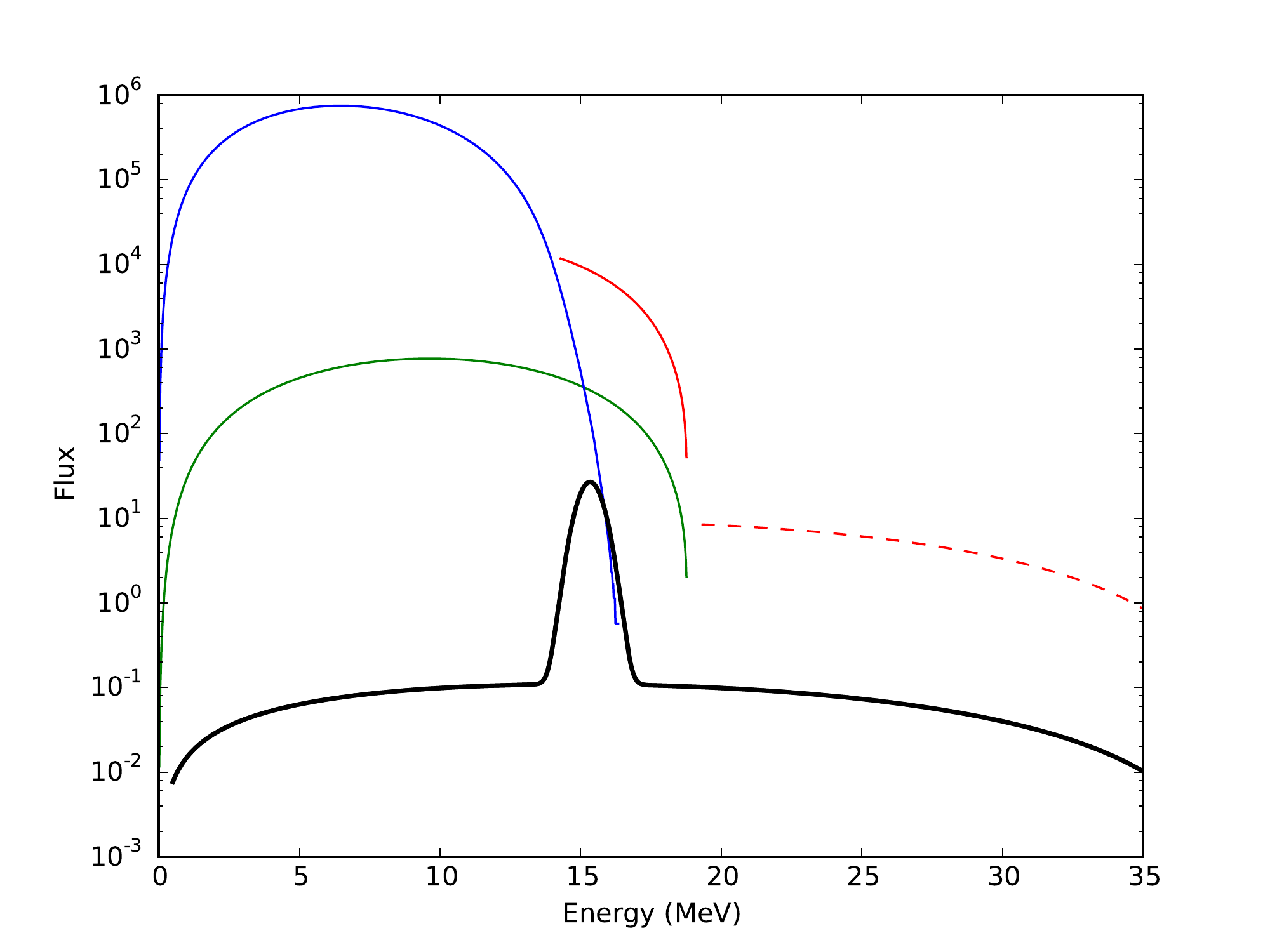}
\caption{Predicted neutrino spectrum (flux in cm$^{-2}$s$^{-1}$MeV$^{-1}$) 
for the NG decays described in the 
text. For comparison the much larger solar neutrino flux from the $^8$B  
(blue) and {\textit{hep}} (green) processes taken from \cite{Bahcall:2004pz} 
is also shown. The red lines 
show present constraints from SuperK \cite{Lunardini:2008xd} 
(dashed) and the SNO {\textit{hep}} search \cite{Aharmim:2006wq} (solid).}
\label{fig:nu_spec}
\end{figure}

The conventional solar neutrino spectrum forms a large background 
to any neutrino signal below $\sim 20$ MeV. These 
solar neutrinos are produced in the rare $^8$B and {\textit{hep}}  ($He+p$) nuclear reactions. The 
$^8$B reactions are well measured and produce a total neutrino flux on the order 
of $6\times10^6$ cm$^{-2}$s$^{-1}$ \cite{Ahmed:2003kj} in good agreement with the 
predictions of the standard solar model \cite{Bahcall:2004pz}. The {\textit{hep}} 
neutrino flux is estimated at $8\times 10^3$cm$^{-2}$s$^{-1}$ \cite{Bahcall:2004pz} 
but has yet to be observationally confirmed with current constraints just slightly 
above the predicted level \cite{Aharmim:2006wq}. Thus, at low energies we should 
compare any nugget associated signal to this known solar background. It should also be 
noted that the uncertainty in both the measured \cite{Ahmed:2003kj} and predicted 
\cite{Bahcall:2004pz} $^8$B spectrum is larger than the flux of neutrinos 
expected to be generated by nugget annihilations. This large background, 
combined with the decreasing sensitivity of observations at low energies, will 
make the detection of any nugget generated neutrino signal challenging. This is 
particularly true given  
that the majority of neutrinos are produced at the $\sim 10$ MeV scale 
through the decay of the lightest NG modes.
The resulting constraints on nugget annihilation rates from the neutrino channel are 
shown in figure (\ref{fig:nu_spec}).

\begin{figure}
\includegraphics[width=\linewidth]{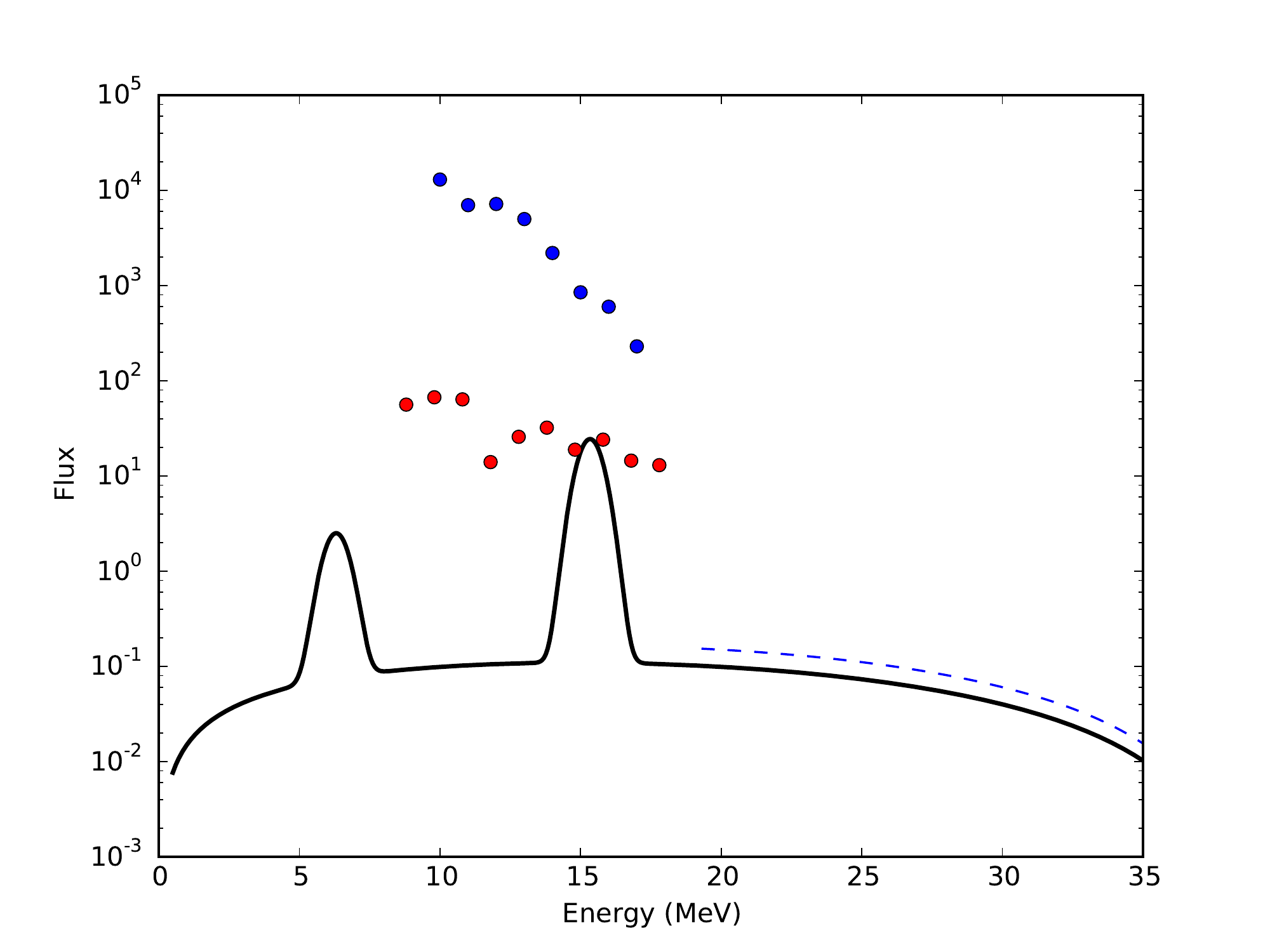}
\caption{Predicted antineutrino spectrum (flux in cm$^{-2}$s$^{-1}$MeV$^{-1}$) 
for the NG decays described in the 
text. Also shown are the high energy SuperK constraints from \cite{Lunardini:2008xd} 
(blue dotted line) and \cite{Gando:2002ub} (blue points) as well as the KamLAND 
constraints from \cite{Collaboration:2011jza} (red points).}
\label{fig:antinu_spec}
\end{figure}

The situation for antineutrinos is somewhat better. In addition to the gain in 
sensitivity from inverse beta scattering mentioned above there is relatively 
little solar background allowing for useful measurements at lower $\sim 10$MeV 
energies. As well as the strong SuperK limits at large energies \cite{Lunardini:2008xd} 
there are constraints from the SuperK solar antineutrino search \cite{Gando:2002ub} 
and well as from KamLAND \cite{Collaboration:2011jza}. The resulting constraints, 
extending down to 8.8MeV are shown in figure (\ref{fig:antinu_spec}).

\exclude{
While the range of CS phases and associated NG mass spectra change from model to model
our basic argument may be illustrated with a neutrino spectra 
for a typical case. Consider a model in which the lightest NG modes have 
masses of 15 MeV and 25 MeV and in which continuum emission is dominated by the 
three body decays from a heavy 80 MeV state. This will result in two decay lines from  
the light modes as described above, and a continuum whose spectrum 
may be estimated from basic kinematics. The resulting neutrino flux 
is show in figure \ref{fig:nu_spec} along with the solar backgrounds and present 
constraints from SNO and SuperK while the (essentially identical) antineutrino 
spectrum is shown in figure \ref{fig:antinu_spec}. In that figure we assume that the solar 
neutrino spectrum is a sum of the standard solar model predictions along 
with an component produced by the annihilation of antiquark nuggets. The constraints are 
then shown as the neutrino spectrum which would produce the maximum allowed integrated 
flux in the 14.3-20MeV and $>$20MeV regions of the SNO, SuperK and KamLAND analysis. }

As may be seen from figures \ref{fig:nu_spec} and \ref{fig:antinu_spec} the 
neutrino flux associated with this representative nugget annihilation model is 
fully consistent with present constraints\footnote{Obviously, had the 
neutrino lines shown in figures \ref{fig:nu_spec} and \ref{fig:antinu_spec}  
been shifted to energies above 20 MeV 
the flux attributed to the nuggets would exceed the constraints set by SuperK in 
that region. This would essentially correspond to the analysis presented in 
\cite{Gorham:2015rfa} where the decay of conventional hadronic states were assumed to 
dominate the nugget annihilation process.}. It should be noted that while, as we have 
argued here, nuggets in a CS phase can generically  evade present constraints relatively 
modest improvements in these constraints (roughly a factor of four) will provide a  
powerful test of the viability of the quark nugget dark matter model. This is a particularly 
strong test of the model because, as shown in \cite{Gorham:2015rfa}, the predicted 
neutrino flux is largely insensitive to the nuggets' mean baryon number across most of the 
applicable range. The solar antineutrino flux may therefore be used to search for a 
quark nugget dark matter signal even if the nuggets are near the higher end of their 
allowed parameter space. This region that is not easily accessible by other means and 
the search may be conducted using currently operating neutrino detectors.

\section*{Conclusion}

The basic claim of the present paper is that the energetic neutrinos required for strong 
constraints from the SuperK data are not produced in the CFL phase and 
this claim also holds for most other CS phases. 
This is because the many other CS phases, such as 2SC (two flavour superconductor) 
all have massless ungapped modes which will dominate the annihilation products of 
interactions with solar matter. In these cases 
a typical neutrino's  energy will be even smaller than in the gapped CFL phase. 
In making this claim we have assumed that the annihilation happens inside the nugget. 

There is a small chance that the annihilation may happen on the sharp surface separating 
the CS from vacuum, in this case the annihilation products could include conventional 
$\pi$ mesons and $\mu$-leptons which indeed produce higher energy neutrinos. 
We think that the probability for this to happen is indeed very small, 
see the comments on this in section \ref{sec:energy-transfer}. For such rare events  
the  analysis of ref.  \cite{Gorham:2015rfa} is fully valid. Unfortunately, it is very hard to 
make any quantitative estimations for this type of annihilation on the surface in strongly 
coupled QCD. Therefore, we follow \cite{Lawson:2010uz} and introduce a 
phenomenological coefficient for such annihilation 
on the  surface $\chi_{\mu}$. This parameter is expected to be very small 
$\chi_{\mu}\ll 1$ because the probability for the annihilation on the interface between the CS 
and vacuum is strongly suppressed.  
With this phenomenological parameter the  interpretation of  
the main results of ref.  \cite{Gorham:2015rfa}  can be restated as follows:  
the anti-nuggets cannot account for more that  $({0.2}/{\chi_{\mu}})$ of the dark matter flux, 
in contrast with original claim in \cite{Gorham:2015rfa}  of a $20\%$ limit.

Finally, we should comment here that similar constraints also hold for   
antiquark nugget annihilation within Earth. We do not elaborate on this subject  
because the parameter space constrained by this process has 
been largely covered by other experiments.

\section*{Acknowledgements} 
 
We are thankful to Peter Gorham for bring to our attention the possible importance of 
neutrino constrains and for subsequent correspondence and comments.  
We also thank an anonymous 
Referee who suggested presenting the antineutrino spectrum separately 
from the neutrino data which resulted in Fig. 
\ref{fig:antinu_spec}. It may bring the $\bar{\nu}$- constraints to the level of 
needed for discovery, as one can see from the plot. 
We also thank Scott Oser who explained to us the basic properties,  features and 
potentials of the existing neutrino detectors,  which considerably improved
our understanding on potentials to detect the $\nu$-signals due to the QN dark matter model.

This research was supported in part by the Natural Sciences and Engineering 
Research Council of Canada.

\end{document}